# Algodoo for Online Education: Impulse and Momentum Activities

**Atakan Çoban**

Yeditepe University, Depertmant of Physics,Yeditepe University, Faculty of Arts and Sciences, Department of Physics, Ataşehir, Istanbul, Turkey.

E-mail: atakancoban39@gmail.com

**Abstract**

During the periods of sudden transition to online education, the opportunity to make applications that might attract students' attention to the course has decreased even more. Although this deficiency was tried to be eliminated with videos and simulations, it was not possible to ensure active participation of students in some cases. In this study, the Algodoo program, which can increase the efficiency of the teaching environment by ensuring active participation of students in online lessons and the applications that can be done about Impulse and momentum are explained in detail. A total of 6 different applications were carried out, 1 related to the subject of impulse, 1 related to the momentum, 2 related to the relationship between impulse and momentum change, and 2 related to momentum conservation. At the same time, while developing these applications, the adjustments made on the simulation and the reasons are explained in detail. In this way, both the introduction of the program and the sample application suggestion were presented. The values obtained as a result of the applications were calculated and compared both theoretically and on simulation in different ways. As a result, it has been observed that the values have internal consistency with each other and are also compatible with theoretical calculations. Algodoo program, which allows many interactive applications and can be downloaded for free, is a program that can be used both in lecturing and evaluation processes in physics lessons while online education process.

## 1. Introduction

Recently, online courses have been conducted almost everywhere in the world[1]. This situation virtualized the lecture environments. Suddenly going into such a process left physics teachers desperate to do in-class activities. When considered specifically in the physics lesson, classroom practices have a very important place. It is very important to organize activities in the online environment as in the classroom[2]. It can be efficient in providing visualization through video or ready-made simulations. However, it may be insufficient to attract the attention of the students and to activate the student in the process.

Within the scope of the physics course, current digital learning environments are divided into two categories as 'constrained' and 'less constrained'[3]. Digital learning environments used in physics education, such as PhET simulations[4], Physlets[5] and QuVis animations[6], are 'constrained'. These are ready-made simulations that focus on the points that experts who prepare simulations in these programs consider the most important. However, since these ready-made simulations allow limited applications, adjustments may not be made on all variables in some subjects. Algodoo[7], Interactive Physics[8], and Fizika[9] programs are 'less constrained' digital learning programs. There are many variables in these programs such as gravity, air drag, flexibility, speed, mass, color, etc. And these variables can be given different values as desired. In this way, various unique simulations that focus on different points within the scope of the relevant subject are designed and provide the opportunity to make original applications. Algodoo interactive physics simulation program, which can be used in classroom applications that require active participation of the student, can be very useful in online lessons[10]. The program is very small and can be downloaded for free[11]. There are many applications on this simulation program, which is developed by considering the laws of physics, especially on Mechanical Physics. Within the scope of the program, there are many possibilities such as adding objects in various geometric shapes, adjusting all kinds of physical properties of the objects, adding the graphics of many variables of the object relative to each other, providing the visualization of the vector representation of the variables of the object, adding an inclined plane at different angles, adding forces in different sizes and directions[12,13].

The implementation of the Algodoo program can be done in several ways:

- First, theoretical explanation is made as chapter by chapter. Then, students are asked to design a simple case study related to the subject on the simulation. In this way, it is ensured that the subject scope is learned by discovering its equivalent in daily life. For example, in order to analyze the relationship between mass and acceleration, forces of the same magnitude are added to objects of different mass, and the relationships between accelerations of objects are analyzed.

- An example problem written using the values of a previously prepared simulation is solved with the class after the lecture. Then the ready-made simulation is opened and the sample real-life problem is analyzed by the students. It is ensured that the accuracy of the result reached by theoretical calculation is tested and the knowledge learned is reinforced. For example, students are asked to calculate the magnitude of the final velocity as a result of the 15 m displacement of a static 2 kg object on a frictionless ground, with a force of 10 N on it. Then, this sample situation is analyzed by providing the necessary visualizations through Algodoo.

- Lectures can be made directly on the algodoo program. First, the situation related to the subject is shown on Algodoo and questions are asked to make students wonder and think about the subject. Then the situation is explained by lecturing and the simulation is taken back and analyzed again, this time with theoretical calculations. For example, by enabling a force to act on an object first, the simulation is stopped after a certain period of time. At this point, it is explained what impulse is and what its equation is. Then the simulation is taken back and the force-time graph of the object is added, the motion is started again and the impulse is calculated.

- Within the scope of an example case, students can be provided to determine the relationship between variables by testing through the program. For example, in an application related to projectile motion, the relationship between the angle and the maximum height can be analyzed by increasing the angle of the initial velocity of the object with the horizontal. Or, in another application, they can analyze the time-dependent graphs of the position and velocity during accelerated motion and how these graphs change depending on the force acting on the object.

In this study, the development process and data analysis processes of 6 sample applications on impulse and momentum that can be used during the lesson using the Algodoo program will be explained.

2. **Theory**

In this section, theoretical information about the applications in the study is explained. The theoretical information used was taken from the relevant textbooks[14]. The product of the net force acting on the object and the duration of action Δt is called impulse and is calculated by the equation

$$\vec{I} = \vec{F}.\Delta t \quad (1)$$

At the same time, the area under the time-dependent graph of the force is equal to impulse. And, objects with a certain mass and speed have momentum. Momentum is calculated by the equation

$$\vec{P} = m.\vec{v} \quad (2)$$

If the net force acting on an object is not zero, the speed of the object changes. The impulse affecting the object causes the momentum of the object to change. The impulse affecting the object is equal to the change in the momentum of the body and this relation is expressed as

$$\vec{I} = \Delta\vec{P} = \vec{P_2} - \vec{P_1} \qquad (3)$$

The total momentum of a system consisting of more than one body is conservative and does not change unless an external force acts on the system. The momentum of the objects in the system can change as a result of various interactions (collision, explosion, etc.) in the system. But the sum of these momenta never changes. In two-dimensional isolated systems, momentum is conservative in both axes, and the final momentum in both axes is equal to the initial momentum in both direction and magnitude.

### 3. Applications & Results

#### 3.1. Impulse

With the help of the Algodoo program, it is possible to develop activities by taking measurements directly on the graph and by putting the data obtained from the simulation into the equations. In order to develop an activity related to impulse using the Algodoo program, first of all, air friction should be removed by clicking on the bottom middle partition. Then, a rectangular object of insignificant dimensions is added with the help of the toolbars at the bottom left. To remove the friction of the object, the "friction" section in the "material" toolbar is set to zero. Then, force is added to the object from the toolbars in the lower left corner. By double-clicking the force, it is possible to adjust the size, the point of action and the activating keys. In this application, the force acts continuously, but it can also be adjusted to press a key to activate the force, if desired. In the study, a force of 5N was preferred. By double clicking on the object, the "show plot" toolbar is selected and the time-dependent graph of the force acting on the x-axis is added. The simulation speed was set as 0.100 by hovering over the start button and the simulation was started. The motion of the object with the added force effect is observed as in figure 1.

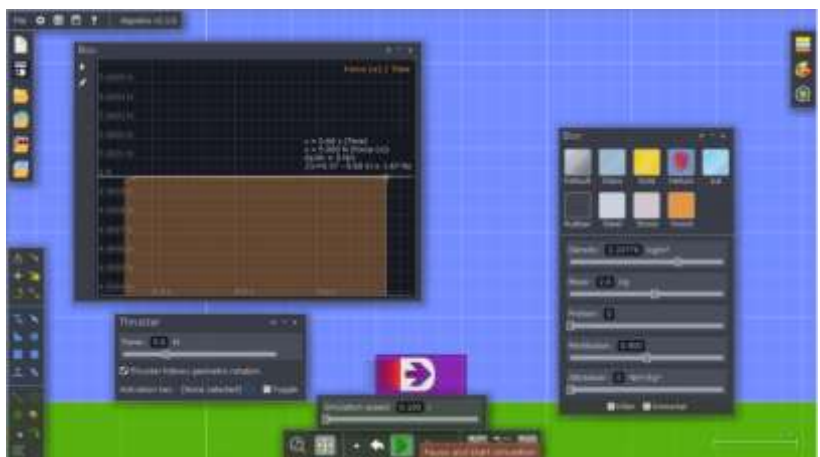

**Figure 1.** Screenshot of the application about Impulse in Algodoo

The simulation was stopped after a while. The added force had an effect from the moment the simulation was started to the moment it was stopped. On the graph, the time that the force is active was determined as Δt = 0.31 s. At this point, the size of the impulse can be calculated as 1.55 N.s with the help of equation 1. The area under the time-dependent graph of the force added to the simulation is also equal to the impulse and this value can be seen directly as 1.67 N.s through the simulation. By clicking on different points on the graphic, the area under the graphic at that time can be determined with the help of the program. There is a high level of agreement between the result obtained in the theoretical calculation made with the values obtained through the simulation and the value read directly on the graph.

### 3.2. Momentum

In the Algodoo program, it is possible to adjust objects in different masses and speeds of different sizes. In addition, velocity magnitudes can be adjusted in both x and y axis. Therefore, it is possible to add objects having momentum in all directions and sizes. With the help of the program, the momentum magnitudes in both axes can be plotted depending on different variables. In Figure 2, a screenshot of the simple application developed for momentum is given.

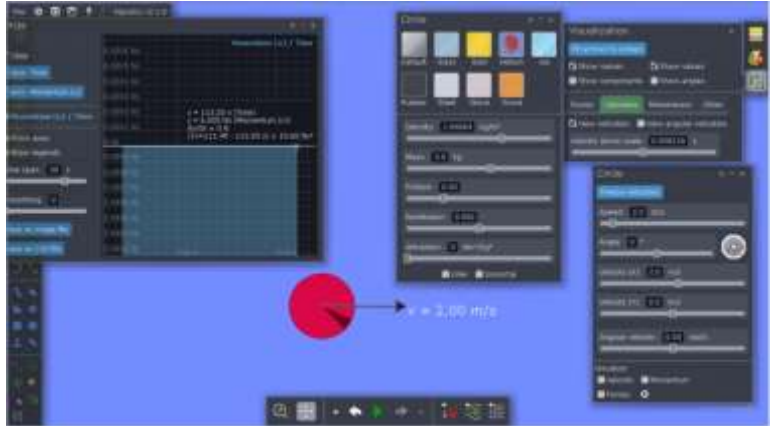

**Figure 2.** Screenshot of the application about Momentum in Algodoo

In the application, air friction and gravity were removed and a circular object was added. By double clicking on the object, the mass of the object is set as 3.0 kg in the "material" section, and its speed is set as 2.0 m / s in the + x direction from the "velocities" section. Time dependent graph of the value of momentum on the x-axis has been added from the "show plot" section. The magnitude of the momentum value of the object on the x-axis was determined as 6.0 N.s both by theoretical calculation and on the graph.

### 3.3. Impulse and momentum change

In order to analyze the relation of impulse being equal to momentum change, a circular body was added in a simulation where air friction and gravity were removed, and a force of 4 N was added on the object. The vector representation and magnitude of the added force are provided with the help of "Visualization" from the toolbars located in the upper right corner of the simulation program. The simulation is finalized as in Figure 3 by adding the time-dependent graph of the momentum of the object on the x-axis.

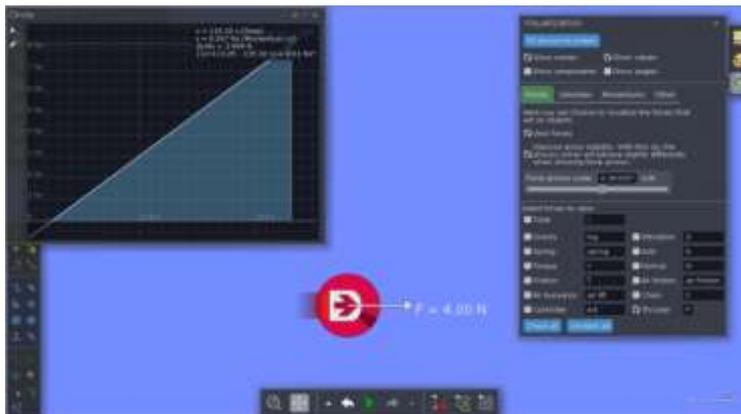

**Figure 3.** Screenshot of the first application about Impulse and Momentum Change in Algodoo

The duration of the force on the graph was determined as 2.07 s and the final value of the momentum was 8.267 N.s. The body was at rest when not under the force, and therefore its momentum was 0. The final momentum of the body is equal to the change of momentum and is 8.267 N.s. During the period of force of 4 N, the size of the impulse was calculated as 8.28 N.s. It is seen that the momentum change value read from the graph added to the simulation and the impulse value calculated using the values in the simulation are equal to each other with a small margin of deviation.

In another application, an object with a mass of 5 kg has been raised to a certain height while standing still. The air friction was removed and the gravitational acceleration was adjusted to be -90º in direction and its magnitude to 9.8 m/s$^2$. Velocity vector is visualized by selecting the view velocity section from the velocities section in visualization from the toolbars in the upper right corner, and the velocity size is also visualized by selecting the show value expression in the same toolbar. By double clicking on the object, a graphic is added as time on the x axis and force (magnitude) on the y axis through the show pilot option. The simulation is started, the movement of the object is monitored and the simulation is stopped. At this moment, the screen shot of the simulation is as in figure 4.

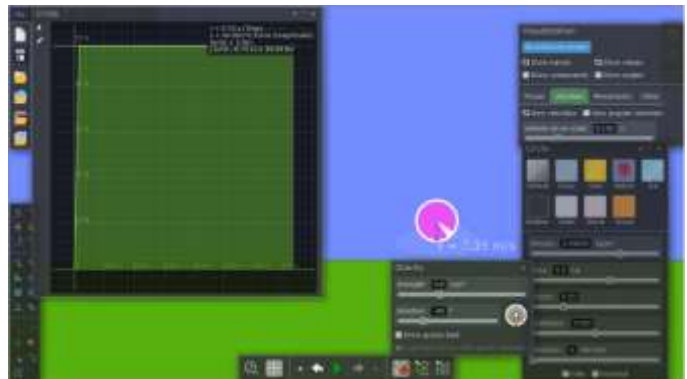

**Figure 4.** Screenshot of the second application about Impulse and Momentum Change in Algodoo

If the gravitational acceleration is assumed to be g = 9.8 m/s$^2$, the net force acting on an object with a mass of 5 kg is the force of gravity and this force is 49 N as can be seen from the graph in Figure 4. The time from the start of the simulation to the moment it is stopped is 0.73 s. Therefore, the impulse affecting the object during this movement is 35.77 N.s. Impulse was found as 35.93 N.s from the area under the force-time graph added to the simulation. The two values are largely compatible with each other. It can be seen from the simulation that the linear velocity of the object that started to move in a stationary state when the simulation stopped is 7.19 m/s. The momentum change of an object with a mass of 5 kg is also ΔP=35.95 N.s. At this point, it is seen that the results achieved have a high internal consistency with each other.

### 3.4. Momentum Conservation

The ability to add objects of different mass and velocities in different directions on the simulation leads to various applications for momentum conservation. In this study, two applications related to two dimensional elastic collision and related to internal explosion have been made. A wide variety of other momentum conservation applications can be developed with the methods used in these applications. In the application related to the collision, two circular objects are added by removing the air friction and gravity and these objects are adjusted to have a mass of 3 kg. The velocities of 4 m/s in the + x direction and 3 m/s in the + y direction are added to the objects as shown in Figure 5.a. For the collision to be fully flexible, the restitution value is set as 0.500 and attraction 0 in the "material" section.

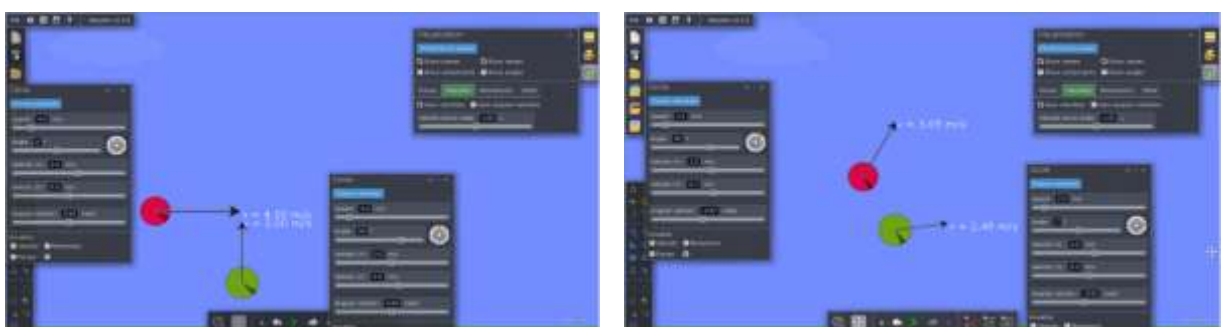

**Figure 5. a)** Screenshot of the initial state of the first application of Momentum conservation in Algodoo **b)** Screenshot of the final state of the first application of Momentum conservation in Algodoo

When Figure 5a. is examined, the first momentum of the red one out of two objects with a mass of 3 kg can be calculated as 12 kg.m/s in the + x direction and 9 kg.m/s in the + y direction of the green one. After the collision, it can be seen that the speed of the red object in the x-axis is 1.5 m/s in the y-axis, 2.7 m/s, the green object in the x-axis 2.5 m/s and 0.3 m/s in the y-axis. Using these values, the final momentum in the x-axis is calculated as 12 kg.m/s and the final momentum in the y-axis is calculated as 9 kg.m/s. These results appear to be the same as the initial momentums in accordance with the conservation of momentum.

In another application, it is related to the explosions that are frequently encountered in momentum conservation. In this application, 3 circular bodies of different masses were nested at the same point while the simulation was in pause. Air friction and gravity have been removed. Visualization of momentum has been selected from the visualization toolbar in the upper right corner. "Show values" and "show components" options are selected in the same section. In this way, the components and sizes of the momentum vectors on the x and y axis are visualized. Simulation was started when the objects were nested at the same point and after a while it was stopped and a view like in figure 6 was obtained.

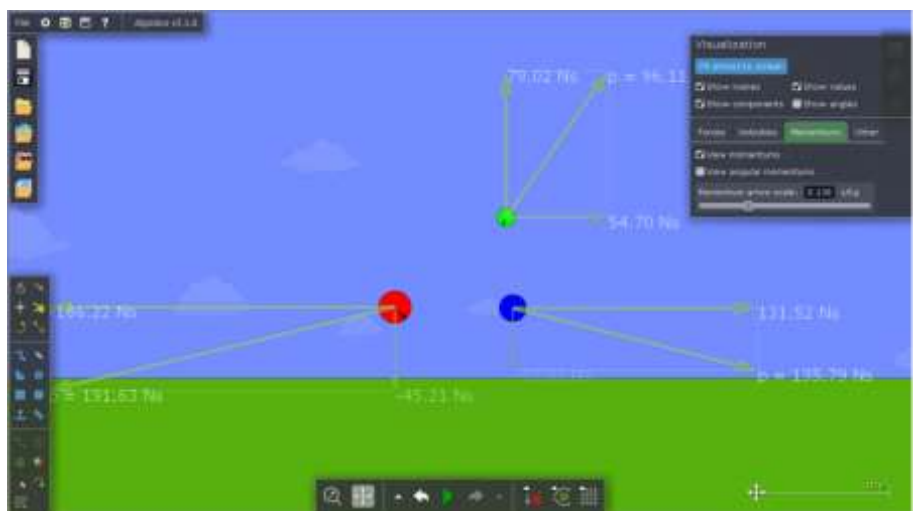

**Figure 6.** Screenshot of the final state of the second application of Momentum conservation in Algodoo

When Figure 6 is examined

- The momentum of a 4 kg red body on the x-axis is -186.22 N.s and -45.21 N.s on the y-axis.

- The blue body with a mass of 3 kg has a momentum of 131.52 N.s on the x-axis and -33.81 N.s on the y-axis.

- The green body with a mass of 2 kg has a momentum of 54.70 N.s on the x axis and 79.02 N.s on the y axis.

Before the simulation started, all three objects were stationary at the same position, and the initial momentum was zero for both the x and y axis. When the momentum values obtained at the moment the simulation is stopped, it is seen that the total final momentum magnitudes on both x and y axes are also zero.

### 4. Conclusion

In online lessons, it is more difficult to attract students' attention to the lesson and to perform efficient educational processes than in the school environment. At this point, it is indispensable to use different methods. Today's students are individuals who grew up in the age of technology and benefit from most of the technology's opportunities. Providing online trainings to such a community with direct instruction will undoubtedly not attract them to actively work in the classroom environment. In order to eliminate this problem and increase the efficiency of the education process, it is necessary to visualize the course environment using videos, animations and simulations. Algodoo program is a free interactive physics simulation program. The program is developed considering the laws of physics, and instead of watching ready-made simulations, optional changes can be made on physical quantities in this program. In this study, sample applications on impulse and momentum are explained. In addition, technical information about the program is also given. Besides these applications, information that will prepare the groundwork for a wide variety of applications on impulse and momentum is given. It is very difficult to make applications due to limited opportunities in these times when the trainings continue online throughout the world, and at this point, applications to be made with the Algodoo program, which can be downloaded and used for free, will be very useful.